\documentclass[pra,twocolumn,superscriptaddress]{revtex4}
\usepackage{graphicx}
\usepackage{dcolumn}
\usepackage{bm}
\usepackage{amsmath}
\usepackage{amsfonts}

\usepackage{fix-cm}

\newcommand{\beq}{\begin{eqnarray}}
\newcommand{\eeq}{\end{eqnarray}}

\newcommand{\eq}[1]{Eq.~(\ref{#1})}
\newcommand{\fig}[1]{Fig.~\ref{#1}}

\begin{document}
\title{Quantum limit of photothermal cooling}

\author{Simone \surname{De Liberato}}
\affiliation{Department of Physics, University of Tokyo, Hongo, Bunkyo-ku, Tokyo 113-0033, Japan}
\author{Neill Lambert}
\affiliation{Advanced Science Institute, RIKEN, Wako-shi, Saitama
351-0198, Japan}
\author{Franco Nori}
\affiliation{Advanced Science Institute, The Institute of Physical and Chemical Research (RIKEN), Saitama 351-0198, Japan}
\affiliation{Physics Department, University of Michigan, Ann Arbor, Michigan, 48109, USA}
\begin{abstract}
We study the problem of cooling a mechanical oscillator using the
photothermal (bolometric) force. Contrary to previous attempts to
model this system, we take into account the noise effects due to
the granular nature of photon absorption. This allows us to tackle
the cooling problem down to the noise dominated regime and to find
reasonable estimates for the lowest achievable phonon occupation
in the cantilever.
\end{abstract}
\maketitle

\section{Introduction}

The optomechanics of deformable cavities was born in the seventies
from the seminal works of Braginsky \cite{braginsky70}.  In recent
years the first successful observation of self-cooling due to
photothermal \cite{metzger04} and radiation pressure
\cite{arcizet06,gigan06,schliesser06} forces started a race to
reach the quantum regime of mechanical motion \cite{aspel09,
cleland10,Roche, Teufel, ash, kwan,
nunnenk,xue,wei1,zhang,ian,mahboob}.

The self-cooling process usually involves an optical cavity whose
mirrors feel an optically-induced force proportional to the cavity
population.  This force depends on the mirror position because the
cavity population depends on the cavity length. If this dependence
is delayed in time, e.g., because the force depends on the
position of the mirror at an earlier time, then self-cooling of
the mirror motion can be achieved (given the right parameter
regimes \cite{favero09}).

These optomechanical self-cooling schemes can be classified into
different categories according to the nature of the
optically-induced force that dominates the cooling process. The
most commonly used is the radiation pressure force that exploits
the pressure exerted by photons bouncing off the mirror. In this
case, the delay in the force is given by the cavity storage time;
that is, the time taken by the photon population inside the cavity
to adjust when the cavity length is modified. The second kind of
optically-induced force exploits the thermal deformation of the
mirror due to the absorption of photons \cite{mertz93}.  In this
case, the delay is given by the heat diffusion time
 through the mirror.  This force, dubbed the photothermal (or
 bolometric)
force, has been observed in multiple experiments and can be much
stronger than the radiation-pressure force in appositely designed
systems \cite{metzger04,metzger08,jourdan08}.

\begin{figure}[]
    \includegraphics[width=\columnwidth]{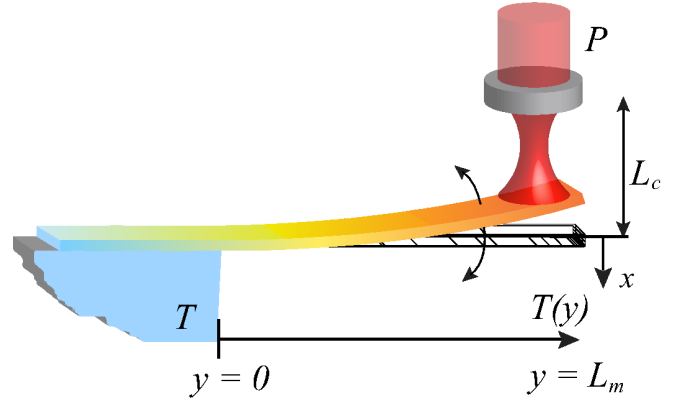}
    \caption{(Color online)
   Schematic diagram of an
optomechanical system.  The mechanical system, or cantilever, is
fixed on the (blue) left edge, and operates as a movable mirror
(right side, in orange) forming one half of an optical cavity
(shown by the red constriction). The grey disk is a non-moveable
(i.e., fixed) half-mirror. The optical and mechanical modes are
coupled via both radiation-pressure and photothermal forces. The
optical cavity (bounded by the fixed grey disk and the (orange)
right edge of the cantilever) has equilibrium frequency
$\omega_c$, quality factor $Q_c$, equilibrium length $L_c$, photon
lifetime $\Gamma_c$, and is pumped by an external laser with power
$P$ and frequency $\omega_p$. The cantilever has thermal
conductivity $\kappa$, length $L_m$, and surface area $s$. The
cantilever mode we consider for cooling has frequency $\omega_m$,
effective mass $m$, and quality factor $Q_m$.  The right edge of
the cantilever absorbs photons at the rate $\alpha$, and has a
thermal deformation coefficient $\chi$.
 \label{sysfig}
 }
\end{figure}

When either cooling mechanism brings the system towards the
quantum regime, it also becomes necessary to consider (together
with the dynamically-induced cooling rate) the spontaneous heating
rate given by the intrinsic quantum noise.  This is because
photons are reflected, or absorbed, one-by-one in both cooling
schemes.  In the case of radiation-pressure-cooling, a quantum
theory exists \cite{law05,wilson07,marquardt07,pater,graj} showing that,
with the ideal parameters, cooling to the ground state of the
mirror motion is still possible.  However, unlike the radiation
pressure force, until now the photothermal cooling process has not
been investigated in the quantum limit. This is due to the
intrinsic difficulty in building a fully quantum theory of a
process that is fundamentally dissipative and involves a
macroscopic number of degrees of freedom (e.g., photon absorption,
heat diffusion, and thermal deformation).

Our goal in this paper is to construct a  theory (albeit a
phenomenological one) rigorous enough to be able to
firmly answer questions of practical importance for experiments that employ
photothermal cooling. Mainly  we want to know whether it is
possible to exploit photothermal cooling to reach the quantum
regime, and to identify the bounds on parameters necessary to reach this regime.

The theory we construct uses the model of a cantilever-mounted
mirror [see \fig{sysfig}] in order to be able to compare with
specific experiments. However, in principle one can also apply our
arguments to other systems, like microdisk optomechanical systems
\cite{armani03,schliesser08}.

This paper is structured as follows: in Section \ref{theory} we lay down our theory and analyze the results.
After having described the issue of residual heating due to photon absorption in Section \ref{residual}, we will test our theory with numerical
data taken from a well documented experiment in Section \ref{numerics}.

\section{Theory of photothermal cooling}
\label{theory}

\subsection{Basic approach}
Our aim is to describe the process of self cooling of an
oscillation mode of a cantilever (see  \fig{sysfig}).  In
the following, unless explicitly stated, we will  always be considering this specific mode.
It is usually the fundamental (lowest frequency) mode of the cantilever, but this depends on
the geometry of the  optical cavity and the cantilever \cite{jourdan08}.

In the specific model we consider here, the cantilever acts as one
of the mirrors of a pumped photonic cavity. The cantilever mode
couples to the photonic field of the cavity because the frequency
of the cavity, and thus its equilibrium population, depend on the
cantilever position. Assuming the pumping laser to be red detuned,
and referring to the coordinate axes as shown in  \fig{sysfig}, we
have that a shift of the cantilever toward positive $x$ will cause
a lowering of the cavity frequency and thus an increase in the
photonic population (and {\it vice versa}).

Even with a very reflective coating there will always be a residual photon absorption in the cantilever. When a photon is absorbed in the cantilever it will create an excess
of heat (and thus an excess phonon population) in the region of the cantilever that serves as the optical cavity mirror. This heat will then diffuse through the cantilever following normal rules of heat transfer.

Anharmonicity in the interatomic  potential in the cantilever gives rise to the phenomenon of thermal deformation  \cite{bauer56}.
As the average energies in the vibrational degrees of freedom of the atoms (and thus the phonon population) increase, the average interatomic distances change. If this deformation is
not uniform through the whole cantilever, (e.g. if the cantilever is formed by layers of materials with different thermal expansion coefficients) then strains develop that exert an effective force on the entire cantilever.

The number of {\em absorbed} photons, and thus the
corresponding excess heat, is proportional to the
{\em total photon population} of the optical cavity. However the deformation-generated-force is exerted only after the heat has diffused through the cantilever.
This implies that the cantilever is subject to a force dependent on its position at past times.

The cavity photons act on the cantilever also through the well
known radiation pressure effect, that is, they exert a force on
the cantilever as they bounce on the movable mirror. In usual
radiation-pressure cooling experiments the cavity response time is
of the order of the cantilever mode frequency (the good cavity regime
\cite{marquardt07,wilson07}), that is, the time the cavity
population takes to adjust to the new cantilever position is of
the same order of the cantilever period. This also results in a
delayed force acting on the cantilever.  However, in the systems
optimized for photothermal cooling which we treat in this paper,
the cavity response time is much shorter than the cantilever
oscillation period (the bad cavity limit) and we can thus
safely neglect the retarded nature of such force.

As mentioned in the introduction, to develop a microscopic quantum theory of the whole process is a formidable
task, due to the intrinsic many-body and dissipative nature of the
phenomena involved (photon absorption and heat transfer).  However, given the time scales of the cantilever motion and of the thermal delay, we can assume that the quantized cantilever mode is evolving under the action of a semi-classical delayed feedback force.
In other words, the whole system (with thermal delay) effectively measures the cantilever position (via the number of absorbed photons), stores it in a classical signal (the heat) and transmits it to a classical actuator (the
cantilever deformation).

The cooling effect thus arises through a semi-classical feedback
or backaction effect, not different in principle from
electronically controlled active classical feedback mechanisms
used in quantum optics \cite {mertz93,yamamoto,milburn,wiseman}.
In order to better understand this parallelism it is worthwhile to
look at the orders of magnitude involved in usual photothermal
experiments. For the cantilever in Ref.~\onlinecite{metzger08} the
cantilever is $220$ $\mu$m long and its thermal diffusion time is
of the order of $0.5$ ms. Comparing with the lifetime of phonons
in silicon, which are in the picosecond scale \cite{nara91}, it is
easy to understand that no quantum coherence survives through the
effective feedback loop.

Here, we thus construct a  Langevin description of the cantilever under the action of the radiation pressure force and of the time-delayed (effective feedback) photothermal force with an added noise source in the effective feedback loop due to the photon absorption shot noise.

\subsection{Treatment of the photothermal force}

It is well known that the photothermal force, in the classical
regime, can be modelled using a time-delayed force, depending on
the position of the cantilever at past times. This approach has
been proven to describe well experiments not only in the cooling
regime \cite{metzger04,metzger08} but also in the regime where
self-sustained oscillations develop \cite{metzger08b}. The force
can thus be written as a time integral over a memory kernel of the
instantaneous force due to the thermal deformation
\begin{eqnarray}
\label{Fph}
F_{\mathrm{ph}}\lbrack x \rbrack(t)=\int_{\mathbb{R}} d\tau \; h(t-\tau)\;
\frac{dF_{\mathrm{def}}(\tau,x(\tau))}{d\tau},
\end{eqnarray}
where the memory kernel associated to the thermal diffusion delay $\tau$ is taken to be
\begin{eqnarray}
\label{h} h(t)=\left[1-\exp{(-t/\tau)}\right]\Theta(t),
\end{eqnarray}
with $\Theta(t)$ the Heaviside function.

From Eqs.~(\ref{Fph}) and (\ref{h}) we see that the instantaneous force $F_{\mathrm{def}}$ coincides with the actual
photothermal force $F_{\mathrm{ph}}$ we would have for a vanishing delay time $\tau$
\begin{eqnarray}
F_{\mathrm{def}}(t,x)\,=\,\lim_{\tau\rightarrow 0}\;
F_{\mathrm{ph}}\lbrack x \rbrack(t).
\end{eqnarray}

As shown in Refs. \onlinecite{shapiro} and \onlinecite{taubman}, in a coherently pumped optical system with a closed feedback loop, it is sufficient to use a semiclassical theory in order to describe the photon absorption shot noise.
We limit ourselves to a semi-classical description of the electromagnetic field in a coherent state. The easiest way to do so is to start from a quantum
Hamiltonian for the  pumped photonic cavity \cite{marquardt07}
\begin{eqnarray}
\label{Hc}
\hat{H}_c=\hbar\omega_c\left(1-\frac{x}{L_c}\right)\hat{a}^{\dagger}\hat{a}+Ee^{i\omega_pt}\hat{a}+Ee^{-i\omega_pt}\hat{a}^{\dagger},
\end{eqnarray}
where $\hat{a}$ is the annihilation operator for a cavity photon, $\omega_c$ is the cavity frequency, $L_c$ its equilibrium
length, $\omega_p$ is the pump frequency, $E$ is a pump term and we assume the loss rate of cavity photon to be given by $\Gamma_c$.
Substituting c-numbers for the photonic operators ($\hat{a}\rightarrow a$), as usual for coherent states, allows us to calculate the population of the cavity.
Given the large discrepancy between the frequency of the cantilever $\omega_m$ (usually in the kHz to MHz range) and of the photonic cavity response time $\Gamma_c$ (usually in the GHz to THz range) we are not
going to consider explicitly the dynamics of the photonic mode, but we will assume that it responds instantaneously to any change in the position of the cantilever.

We thus abtain the  instantaneous photon population of the cavity as a function of the cantilever position $x$,
\begin{eqnarray}
\label{nc}
n_c(x)&=&\frac{E^2}{\lbrack\omega_c(1-\frac{x}{L_c})-\omega_p\rbrack^2+\Gamma^2_c/4}.
\end{eqnarray}
Supposing that we have a well defined cavity frequency $\omega_c$,
we can find the relation between the pump term $E$ and the input
power $P$ by equating the number of incoming photons $P/\hbar
\omega_p$ with the number of photons lost $n_c\Gamma_c$. We thus
obtain
\begin{eqnarray}
E=\sqrt{\frac{\Gamma_c P}{4\hbar\omega_p}},
\end{eqnarray}
where $\Gamma_c$ is the cavity photon lifetime.

In order to proceed in the derivation of a quantum Langevin equation for the cantilever, which takes into consideration photon shot noise, we need to fix the dependency of the photothermal force upon the intensity of the cavity field $n_c(x)$.

While the microscopic derivation of such a force would be
extremely complex and sample-dependent, we expect it to be a
function of the heat absorbed by the cantilever.  In other words,
the instantaneous photon energy $\omega_c(1-\frac{x}{L})$
times the current of absorbed photons $I_c(x,t)$. Here we
will limit ourselves to the simple case of a linear dependence
that, as we will see, gives good results when compared with
experiments. We will thus write the instantaneous deformation
force as,
\begin{eqnarray}
\label{Fdef} F_{\mathrm{def}}(t,x)=\chi \hbar
\omega_c\left(1-\frac{x}{L_c}\right)I_c(x,t),
\end{eqnarray}
where $\chi$ is a phenomenological deformation coefficient of the cantilever with
dimension of the inverse of a velocity.

We note that the current of absorbed photons $I_c$ can be written
as an average component and a fluctuating component due to shot
noise, the average being proportional to the number of photons
present in the cavity (itself a function of $x$) and the noise to
a coefficient times a stochastic noise term $\eta(t)$
\cite{shapiro,taubman}
\begin{eqnarray}
\label{Ic}
I_c(x,t)&=&\alpha n_c(x)+\delta I_c(t,x)\\&=&\alpha n_c(x)+N(x)\eta(t),\nonumber
\end{eqnarray}
where $\alpha$ is the absorption rate of the cantilever.

In order to have a cooling effect the noise term in \eq{Ic} needs
to be small compared with the average value. Assuming the
cantilever oscillations to be much smaller than the equilibrium
cavity length $L_c$, we can thus insert \eq{Ic} into \eq{Fdef} and
make a first order expansion both $x$ and $\eta$. We thus obtain
\begin{eqnarray}
\label{Fdef2}
F_{\mathrm{def}}(t,x)&\simeq&F_{0}+(\nabla F) x +N \; \eta(t).
\end{eqnarray}

As the different absorption events are completely uncorrelated we
can, in a time interval short enough for neglecting the variation
of $x$, consider the number of absorptions as a Poisson process.
We thus define \beq n_p(t)=\int_0^t dt' I_c(x,t'),\eeq as the
number of photons absorbed up to time $t$. Its momenta are given
by
\begin{eqnarray}
\langle n_p(t) \rangle&=&t\, \alpha\, n_c(x),\\
\langle n_p(t)^2 \rangle&=&t^2 \alpha^2 n_c(x)^2+N^2 \int_0^t\!\!
dt' \int_0^t\!\! dt''\! \langle \eta(t') \eta(t'') \rangle.
\nonumber
\end{eqnarray}
Imposing that $n_p$ has Poissonian, time independent
statistics over the time interval we consider here, we have
\begin{eqnarray}
\langle n_p(t) \rangle&=&\langle n_p(t)^2 \rangle-\langle n_p(t)
\rangle^2.
\end{eqnarray}
We thus arrive at the following condition for the correlator of current fluctuation in the linear approximation
 \begin{eqnarray}
N^2 \langle\eta(t')\eta(t'')  \rangle&=&
\alpha n_c(0)\delta(t'-t'').
\end{eqnarray}
The current $I_c$ is thus characterized by a white noise spectrum
(as in the semi-classical treatment of optical feedback \cite{shapiro,taubman})
 and can be written as
\begin{eqnarray}
\label{Ic2} I_c(x,t)&=&\alpha
n_c(x)+\sqrt{\alpha n_c(0)}\;
\eta(t),
\end{eqnarray}
where the white noise term $\eta(t)$ has the correlator
 \begin{eqnarray}
\langle \eta(t')\eta(t'') \rangle&=&\delta(t'-t'').
\end{eqnarray}

Putting together Eqs.~(\ref{Ic2}),(\ref{nc}) and (\ref{Fdef}) and applying basic analysis we find the following values for the coefficients $\nabla F$ and $N$
\begin{eqnarray}
\label{original}
\nabla F&=&\frac{n_c(0) \alpha \chi \hbar \omega_c}{L_c}\frac{2\omega_c \Delta -\Delta^2-\frac{\Gamma_c^2}{4}}
{ \Delta^2+\frac{\Gamma_c^2}{4}},\\
N&=&\chi \hbar\omega_c\sqrt{\alpha n_c(0)},\nonumber
\end{eqnarray}
where $\Delta=\omega_c-\omega_p$ is the cavity detuning.  In
section IID we use these terms directly in the equation of motion
for the cantilever.

\subsection{Treatment of the radiation pressure force}

Radiation pressure force lends itself  to a microscopic quantum
treatment and it has been studied in various publications
\cite{marquardt07, wilson07}. These works show that the effect of
radiation pressure cooling can be easily modelled by an effective
damping rate $\Gamma_{\mathrm{rp}}$ and an associated noise term
$F_{\mathrm{rp}}$.

The amplitude of $\Gamma_{\mathrm{rp}}$, that is the cooling
capacity of the radiation pressure force depends on the ratio
between the cavity damping and the cantilever frequency and thus
becomes negligible in the bad cavity regime we consider here.  The
same is not true for the noise term, that instead only depends on
the intrinsic magnitude of the force and thus leads to a nonnegligible equilibrium noise population.

From Ref. \onlinecite{marquardt07} we find the following formulas for the optical damping term and for the equilibrium noise population $n_m^{\mathrm{rp}}$
\begin{eqnarray}
\label{rpnoise}
\Gamma_{\mathrm{rp}}&=&\frac{4n_c(0)\hbar \Gamma_c \omega_c^2 }{mL_c^2}\frac{\Delta}{(\Delta^2+\frac{\Gamma_c^2}{4})^2},\nonumber \\
n_m^{\mathrm{rp}}&=&\frac{\Delta^2+\frac{\Gamma_c^2}{4}}{4\omega_m\Delta}.
\end{eqnarray}
Having the damping rate $\Gamma_{\mathrm{rp}}$ and the noise
population $n_m^{\mathrm{rp}}$ in this form, and because we have
assumed the bad-cavity limit, means that we do not have to
explicitly consider the dynamics of the optical cavity.

\subsection{System dynamics: feedback in a quantum Langevin equation}

Using the results from the previous sections, we can describe the
dynamics of the system with a set of coupled Langevin equations
describing the cantilever mode coupled to a thermal bath and
evolving under the conjunct effect of the photothermal and the
radiation pressure force
\begin{eqnarray}
\label{langevin}
\dot{x}&=&\frac{p}{m},\\
\dot{p}&=&-m\omega_m^2x-
(\Gamma_m+\Gamma_{\mathrm{rp}}) p+F_{\mathrm{th}}+F_{\mathrm{ph}}\lbrack
x\rbrack +F_{\mathrm{rp}}\nonumber,
\end{eqnarray}
where $\Gamma_m$ and $F_{\mathrm{th}}$ are the dissipation and the fluctuation terms given by the coupling with the bath \cite{gardinerbook,openquant,ciuti06}.

Using the expression for  the linearized time-delayed force from
\eq{Fdef2} in \eq{langevin} we obtain the following second-order
dynamical equation
\begin{eqnarray}
\ddot{x}(t)+\omega_m^2x(t)+(\Gamma_m+\Gamma_{\mathrm{rp}}) \dot{x}(t)=\frac{F_{\mathrm{th}}(t)}{m}+\frac{F_{\mathrm{rp}}(t)}{m}\\
+ \int_{\mathbb{R}} dt' h(t-t') \left[ \frac{\nabla F}{m}
\frac{d}{dt'} x(t') + \frac{N}{m} \frac{d}{dt'}\eta(t') \right].
\nonumber
\end{eqnarray}
This is an integro-differential equation, and the integral has a
convolution. It thus becomes an algebraic equation in Fourier
space
\begin{eqnarray}
\label{langevininfourier}
&&\left[ \omega_m^2-\omega^2+i\omega(\Gamma_m+\Gamma_{\mathrm{rp}})- \frac{\nabla F}{m(1+i\omega\tau)}\right] x(\omega)=\nonumber \\
&&\frac{F_{\mathrm{th}}(\omega)}{m}+\frac{F_{\mathrm{rp}}(\omega)}{m}+\frac{N\eta(\omega)}{m(1+i\omega\tau)}.
\end{eqnarray}

As we see from \eq{langevininfourier}, the photothermal force
results in a frequency-dependent damping and a renormalization of
the cantilever frequency (see also the discussion in Ref.
\onlinecite{metzger08}). In the usual (rigid cantilever) regime in
which the frequency of the cantilever, renormalized by the
constant ($\omega=0$) part of the photothermal force,
\begin{eqnarray}
\label{WM}
\tilde{\omega}_m=\sqrt{\omega_m^2-\frac{\nabla F}{m}},
\end{eqnarray}
does not differ much from $\omega_m$, the essential part of the
dynamics takes place at $\omega=\omega_m$, and we can thus take
the photothermal-induced damping to be
\begin{eqnarray}
\label{gammaph} \Gamma_{\mathrm{ph}}=\frac{\tau \nabla
F}{m(1+\tau^2 \omega_m^2)}.
\end{eqnarray}
Notice that,  as explained in Ref.~\onlinecite{metzger08}, the
value at which \eq{WM} vanishes corresponds to the onset of mirror
instability and sets an upper limit to the strength of the
photothermal force
\begin{eqnarray}
\label{upper}
\frac{\nabla F}{m\omega_m^2}<1.
\end{eqnarray}

Neglecting quantum (zero-point) fluctuations in the photothermal
force we can thus calculate the phonon equilibrium population due
to the photothermal force alone (i.e., neglecting both the
coupling to the thermal bath and the radiation pressure force in
\eq{langevininfourier}) as
\begin{eqnarray}
n_m^{\mathrm{ph}}=\frac{N^2}{2\hbar \omega_m \tau \nabla F}.
\end{eqnarray}

In principle we can also derive the total equilibrium phononic
population due to the joint effect of the thermal bath, the
radiation pressure and the photothermal effect by integrating
\eq{langevininfourier} (under the rigid-cantilever approximation).
However it is easier to calculate this total population via a
simple rate equation, which gives
\begin{eqnarray}
\label{tot} n_m^{\mathrm{tot}}=\frac{\Gamma_m
n_m^{\mathrm{th}}+\Gamma_{\mathrm{rp}}
n_m^{\mathrm{rp}}+\Gamma_{\mathrm{ph}} n_m^{\mathrm{ph}}}
{\Gamma_m+\Gamma_{\mathrm{rp}}+\Gamma_{\mathrm{ph}}},
\end{eqnarray}
where $n_m^{\mathrm{th}}$ is the equilibrium population at
temperature $T$ that, given the low frequencies of the cantilever
modes, can be approximated as $kT/\hbar \tilde{\omega}_m$.

Inserting Eqs.~(\ref{original}), (\ref{rpnoise}) and
(\ref{gammaph}), into \eq{tot} and remembering that we are working
in a regime where $\Gamma_{\mathrm{rp}}\ll \Gamma_m \ll
\Gamma_{\mathrm{ph}}$, we can rewrite \eq{tot} as
\begin{eqnarray}
\label{totdev}
n_m^{\mathrm{tot}}&=&n_m^C+n_m^N,
\end{eqnarray}
where
\begin{eqnarray}
\label{classical}
n_m^C&=&\frac{kT m\omega_m^2}{\hbar \tilde{\omega}_mQ_m \nabla F}\frac{1+\omega_m^2 \tau^2}{\omega_m \tau}
\end{eqnarray}
is the {\it classical} population due to photothermal cooling and
\begin{eqnarray}
\label{noise}
n_m^N&=& \frac{ \Gamma_c}{\alpha}\frac{1}{ \chi \omega_c L_c}\frac{1+\omega_m^2 \tau^2}{\omega_m \tau}\frac{\omega_c^2}{2\omega_c\Delta-\Delta^2-\frac{\Gamma_c^2}{4}}\nonumber\\&+&
 \chi\omega_c L_c\frac{1}{2\omega_m \tau}\frac{\Delta^2+\frac{\Gamma_c^2}{4}}{2\omega_c\Delta-\Delta^2-\frac{\Gamma_c^2}{4}}
 \end{eqnarray}
 is the population due to noise effects.  The first line in
 \eq{noise} stems from the radiation pressure noise contribution, and the
 second from the photothermal noise contribution.

\subsection{Analysis of the results: the quantum regime}

From \eq{classical} we can, using \eq{upper}  and optimizing over
$\nabla F$, find (consistent with previously known results
\cite{metzger04,metzger08}) that the efficiency of classical
photothermal cooling is maximal for $\omega_m\tau=1$ and its
ultimate value depends on the quality factor of the cantilever
\beq Q_m=\frac{\omega_m}{\Gamma_m}.\eeq The optimization of
\eq{classical} gives us a {\em lower bound} on the phonon
population
\begin{eqnarray}
\label{boundclassical}
n_m^{C,\mathrm{min}}&\geq&\frac{3\sqrt{3}kT}{\hbar \omega_mQ_m}.
\end{eqnarray}

Thus our estimate for the minimal classical population for the
cantilever is proportional to its thermal, equilibrium occupation
$kT/\hbar {\omega}_m$ divided by the quality factor of the
mechanical oscillator $Q_m$. Given the rather low frequencies of
the mechanical modes involved, the condition of having an average
occupation number lower than one puts a rather harsh requirement
on the  mechanical quality factor. In Table I we compare some data
taken from experimentally realized cantilevers. We see that for
the stressed silicon cantilever studied in
Ref.~\onlinecite{verbridge08} the theory predicts, assuming
optimal parameters and liquid Nitrogen temperatures, a thermal
population almost in the quantum regime. Note that only the
cantilever in Ref.~\onlinecite{metzger08} has actually been used
for photothermal cooling, and thus it is the only cantilever for
which we can estimate the deformation coefficient $\chi$ and the
noise contribution to the final population (see Section
\ref{numerics}).
\begin{table}
\begin{tabular}{|c|c|c|c|c|c|}  \hline Reference &  $L_m$ & $\omega_m$ & $Q_m$ & $n_m^{\mathrm{th}}$ & $n_m^{C,\mathrm{min}}$, \eq{boundclassical} \\
\hline
Ref.~\onlinecite{verbridge08} & $275$ $\mu$m & $6.5$ MHz & $1.5\times 10^6$ & $1.2\times 10^6$ & $5$ \\
Ref.~\onlinecite{metzger08} & $220$ $\mu$m & $46$ KHz & $2.2\times 10^3$ & $1.7\times 10^8$ & $6.4\times 10^5$ \\
Ref.~\onlinecite{favero07} & $3.9$ $\mu$m & $3.4$ MHz & $2.9\times 10^3$ & $2.2\times 10^6$ & $7.8\times 10^3$ \\
\hline
\end{tabular}
\label{table1} \caption{Length $L_m$, frequency $\omega_m$,
quality factor $Q_m$, thermal equilibrium population
$n_m^{\mathrm{th}}$ at $77$ K and final population
$n_m^{C,\mathrm{min}}$ from \eq{boundclassical} for different
cantilevers reported in the literature.}
\end{table}

Assuming that we have a cantilever quality factor $Q_m$ large
enough to bring $n_m^{C}$ into the quantum regime, we are thus
confronted with the noise contributions in \eq{noise}. The noise
in \eq{noise} depends on various parameters, and the two noise
contributions (the radiation pressure in the first line and the
photothermal in the second) often have inverse dependencies upon
the parameters.  Thus naively minimizing one noise term can
enlarge the other. We are thus obliged to carefully optimize the
parameters in order to calculate the minimal noise population. We
start by rewriting \eq{noise} as
\begin{eqnarray}
\label{step1} n_m^N&=&\left[\frac{ \Gamma_c}{\alpha}
\frac{1+\omega_m^2 \tau^2}{\omega_m^2 \tau^2}Q_c^2A+
\frac{1}{2A}(\tilde{\Delta}^2+\frac{1}{4})\right] \\ &\times&
\left[2Q_c\tilde{\Delta}-\tilde{\Delta}^2-\frac{1}{4}\right]^{-1},\nonumber
 \end{eqnarray}
 where we have defined \beq A=\frac{\omega_m \tau}{\chi \omega_c L_c},\eeq and introduced the renormalized detuning \beq \tilde{\Delta}=\Delta/\Gamma_c, \eeq and the cavity quality factor
  \beq Q_c=\omega_c/\Gamma_c. \eeq  Again the first term in the numerator of \eq{step1} is from the radiation pressure noise, and the second is from the photothermal noise.  Thus, the new parameter $A$
 encapsulates the way in which the two noise terms are inversely
 proportional to one another.
 Optimizing \eq{step1} over $A$ we obtain
\begin{eqnarray}
\label{step2}
n_m^N&=& \sqrt{2\frac{ \Gamma_c}{\alpha} \frac{1+\omega_m^2 \tau^2}{\omega_m^2 \tau^2}}
\frac{Q_c\sqrt{\tilde{\Delta}^2+\frac{1}{4}}
}{2Q_c\tilde{\Delta}-\tilde{\Delta}^2-\frac{1}{4}},
 \end{eqnarray}
 where the optimal value of $A$ is given by
\begin{eqnarray}
\label{aopt}
A_{\mathrm{opt}}&=&\sqrt{\left(\tilde{\Delta}^2+\frac{1}{4}\right)
 \frac{\alpha}{ 2\Gamma_c Q_c^2} \frac{\omega_m^2 \tau^2}{1+\omega_m^2 \tau^2}}.
 \end{eqnarray}
This optimal value of $A$ makes the two noise terms equal, and
minimum.  The second factor in \eq{step2} can be further optimized
over $Q_c$, yielding its minimum for large $Q_c$
\begin{eqnarray}
\label{step3}
n_m^N&=& \sqrt{\frac{ \Gamma_c}{\alpha} \frac{1+\omega_m^2 \tau^2}{2\omega_m^2 \tau^2}}
\sqrt{  1+\frac{1}{4\tilde{\Delta}}}.
 \end{eqnarray}
 Choosing a large enough detuning $\tilde{\Delta}$ we can thus ignore the second factor in \eq{step3} and we obtain our final estimate for the minimum noise population
\begin{eqnarray}
\label{step4}
n_m^{N,\mathrm{min}}&\geq& \sqrt{\frac{ \Gamma_c}{\alpha} \frac{1+\omega_m^2 \tau^2}{2\omega_m^2 \tau^2}}.
 \end{eqnarray}
We see that, given a large enough cavity quality factor $Q_c$, and
the optimal value for $\chi$ from \eq{aopt}, two factors determine
the minimum noise population: the delay parameter $\omega_m \tau$
and $\Gamma_c/\alpha$. The second one, $\Gamma_c/\alpha$, is the
ratio between the total photon loss rate ($\Gamma_c$, the sum of
the mirror absorbtion $\alpha$ and other radiative losses), and
the loss part due to absorption in the moving mirror alone (thus
this ratio is always larger than one). Equation (\ref{step4})
tells us that, due to the interplay between the radiation pressure
and photothermal noises, while it is possible to reach the
``quantum regime'' with appropriate parameters, it is not possible
to get arbitrarily close to the ground state as in the case of
radiation-pressure-dominated experiments \cite{marquardt07}. In
order to reach the quantum regime it is necessary not only to
limit all the losses other than the cantilever absorption in order
to lower $\Gamma_c/\alpha$, but also to design samples with rather
large values of $\omega_m \tau$. Furthermore, while large values
of $\omega_m \tau$ help lower the noise population, \eq{classical}
tells us that they correspondingly reduce the classical cooling
efficiency, thus increasing the classical phonon population.

To reiterate, to reach the quantum regime we require the following
conditions: \begin{itemize} \item The feedback cooling term,
\eq{boundclassical}, tells us that we need a large cantilever
quality factor $Q_m$, large cantilever frequency $\omega_m$, and
low initial temperature $T$, so that $Q_m \gg kT/\hbar \omega_m$.
\item To simultaneously minimize the photothermal and radiation
pressure noises, we need a high quality optical cavity $Q_c\gg 1$,
a large enough detuning $\Delta$, and a deformation constant
$\chi$ given by \eq{aopt}. \item These conditions give us
\eq{step4}, the final equilibrium phonon population due to noise,
which is then minimized by choosing $\Gamma_c/\alpha$ small.
Again, this can further be minimized by increasing $\omega_m
\tau$, at the cost of reducing the feedback cooling efficiency,
and thus requiring a higher cantilever quality factor $Q_m$.
\end{itemize}

\section{Residual heating from photon absorption}
\label{residual}
Since the photothermal force is due to photon absorption,
the same process that gives the cooling is simultaneously heating
up the cantilever as a whole. This is a rather intriguing aspect of
photothermal cooling that we have, until now, neglected.

This heating mechanism does not modify the fundamental bounds we derived, accounting simply for an increased temperature in \eq{classical}
However, the photothermal heating effect is  important if we want to quantitatively fit our model with experimental data, as it can noticeably influence the observed temperature.

The exact description of such residual heating on the cantilever
motion would depend on the microscopic structure of the cantilever
and it is thus not  in the scope of this paper. Here we will limit
ourselves to a rather phenomenological model, able to mimic the
interesting physics, and depending on only one adjustable
parameter. The simplest way to take into account this phenomenon
is to use, inside \eq{classical}, not the environmental
temperature $T$, but an average equilibrium temperature under
constant illumination (average because the temperature profile
will in general vary over the mechanical mode region).

We now consider a rectangular cantilever, characterized by a
constant thermal conductivity (constant in space and in
temperature), ignoring radiative processes and assuming that the
absorption processes happen homogeneously over one of the surfaces
(see \fig{sysfig} for a schematic illustration). Calling $s$ the
cantilever surface, $L_m$ its length and $\kappa$ its thermal
conductivity, from our approximations and Fourier's law we obtain
a linear temperature profile,
\begin{eqnarray}
T(y)&=&T+\frac{\alpha n_c(0) \hbar \omega_c y}{s\kappa}.
\end{eqnarray}
The effective temperature to be used in \eq{classical} will thus be an average between $T(0)$ and
$T(L_m)$, taking into account that the mechanical mode profile is
not distributed uniformly over $y$
\begin{eqnarray}
\label{epsilon}
\bar{T}&=&T+\frac{\alpha n_c(0) \hbar \omega_c L_m}{\epsilon s\kappa},
\end{eqnarray}
where the adjustable parameter $\epsilon$ is of the order of unity
($\epsilon=2$ for an arithmetic mean).  We include this effect in
our estimates of $\chi$ in the next section.

\section{Existing experiments on photothermal cooling: estimating $\chi$}
\label{numerics} In order to test our model and to gain some
insight in the value of the key parameter $\chi$, we analyze the
results from Ref. \onlinecite{metzger08} that, to the best of our
knowledge, is the only publication reporting the observation of
photothermal cooling with enough experimental data to allow a
comparison with our theory. In Ref.~\onlinecite{metzger08} the
temperature of the cantilever mode is measured for different laser
powers, and for temperatures between $300$ K and $32$ K. All the
material parameters but $\frac{\Gamma_c}{\alpha}$ can be retrieved
directly from the aforementioned reference.

Fitting the temperature for each different laser power allows us
to fix an estimate for the value of $\chi$ around the value $2
\times 10^{-5}$ s m$^{-1}$. With these parameters, we obtain a
phonon noise population of $1.4\times 10^4$, that is a phonon
temperature of $5$ mK.

 \section{Conclusions}

In summary, we have extended the classical model of photothermal
cooling by Metzger et al.\cite{metzger04} to take into account the
effect of quantum noise on cooling efficiency. Rather than build
up a quantum model from a microscopic Hamiltonian, we treated the
complex many-body thermal cantilever deformation effect as an
effective time-retarded force.    As one cools the mechanical
motion to the ground state, noise due to both photon absorbtion
and radiation pressure begins to play a role, and we explicitly
included these, and heating due to absorbtion, in our model.

In the future it will be interesting to test the predictive power of our model against
a larger experimental data-set than the one currently available.

\section{Acknowledgments}
We thank I. Favero and coworkers, who are conducting parrallel work on
the same topic as the one treated in this paper \cite{restepo}, for useful discussions
and M. Ueda, S. M. Girvin, J. R. Johansson, K. Maruyama and T. Brandes for their support.  SDL
acknowledges FY2009 JSPS Postdoctoral Fellowship for Foreign
Researchers.   NL is supported by the RIKEN FPR program.  FN
acknowledges partial support from the Laboratory of Physical
Sciences, National Security Agency, Army Research Office, Defense
Advanced Research Projects Agency, Air Force Office of Scientific
Research, National Science Foundation Grant No. 0726909, JSPS-RFBR
Contract No. 09-02-92114, Grant-in-Aid for Scientific Research
(S), MEXT Kakenhi on Quantum Cybernetics, and Funding Program for
Innovative R\&D on S\&T (FIRST).

\end{document}